# Calorimetry for ILC Experiments: CALICE Collaboration R&D

D.S.Bailey (for the CALICE collaboration)
*University of Manchester, Oxford Road, Manchester, M13 9PL, UK.*

The CALICE Collaboration is carrying out R&D into calorimetry for a detector at the International Linear Collider (ILC). The detector is designed to reconstruct individual particles in dense jets in order to achieve the optimal jet energy resolution for ILC physics using a technique known as Particle Flow (PFA). This imposes significant challenges on calorimeter designs, requiring them to be "imaging" devices – highly granular in order to separate the showers from individual particles. CALICE is investigating a range of technologies, suitable for PFA, for both electromagnetic and hadronic calorimetry. An overview of the prototypes and selected test-beam results are presented.

## 1. INTRODUCTION

ILC physics depends crucially on the ability of the detectors to reconstruct the $e^+e^-$ final states with hitherto unprecedented accuracy. In order to fully exploit the luminosity delivered by the machine it is essential that final states involving many hadronic jets are optimally reconstructed, implying that the detector resolution does not have a significant effect on the measurement of the natural widths of, for example, the $W$ and $Z$ gauge bosons. The Particle Flow approach to reconstruction promises to achieve this goal, which in practice means measured jet energy resolutions of around $30\%/\sqrt{E}$ at the $Z$ mass. In order to achieve this goal, calorimeters must be constructed that are capable of unambiguously identifying the showers from individual particles, which leads to the idea of "imaging calorimeters". CALICE is developing these devices, using a wide range of technologies that are suitable for both electromagnetic and hadronic calorimetry using either analogue or digital readout. For the first time, the readout electronics will be integrated into the detectors themselves in order to minimise the insensitive regions of the calorimeters.

## 2. ELECTROMAGNETIC CALORIMETRY

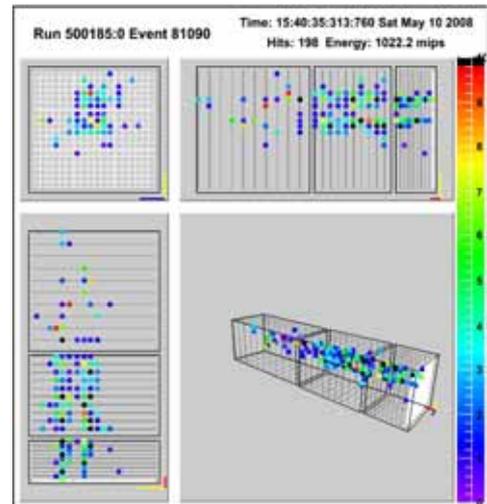

Figure 1: A 2-electron event in the silicon-tungsten ECAL prototype

Two options for the electromagnetic calorimeter (ECAL) are being developed. Both use tungsten for the absorber material as it produces very compact electromagnetic showers, but differ in the choice of active sensor material.

### 2.1. Silicon Tungsten

This option uses silicon diode pads as the active material. The current prototype has 30 $1 \times 1$ cm$^2$ diode layers which instrument the tungsten aborbers (which are either 1.4, 2.8 or 4.2 mm thick). The prototype has a total of 9720 channels. The pads are read-out by ASICs which both digitise and buffer the signals before transferring them to the DAQ. Eventually, these readout chips will be integrated into the detector itself, but in the prototype they are mounted on a PCB which is external to the active detector volume.

The prototype has been exposed to electron beams of various energies at both DESY and CERN. An event with two electrons incident in the prototype is shown in figure 1. The showers of both particles can be seen in the calorimeter. The challenge for the particle flow algorithms is to unambiguously resolve the showers, so the calorimeter is optimised for shower separation and pattern recognition. The energy resolution of the prototype for single electrons is shown in figure 2. Its linearity is better than 1% with a stochastic term of $16.7\%/\sqrt{E}$.

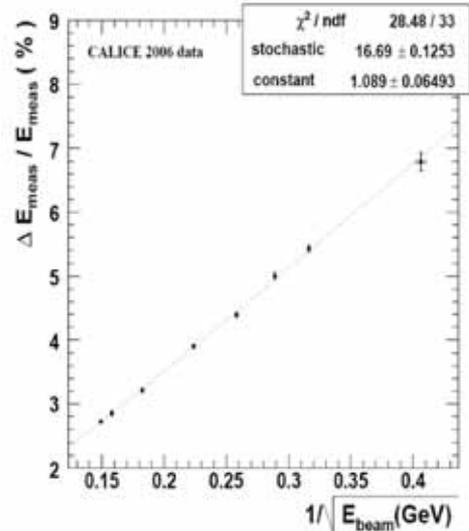

Figure 2: Energy resolution of the Silicon-Tungsten ECAL prototype.

### 2.1.1. Digital ECAL option

Since the goal of the ECAL is to uniquely identify individual particle showers, maximizing the granularity of the sensitive material could improve shower separation in dense jets. One way to do this is to use CMOS pixels (MAPS) instead of the silicon diodes used in the current prototype. Such pixel-based systems could simply replace the silicon diodes using the same mechanical form-factor, and have the potential to be very cheap to produce in large quantities since they use a standard industrial process. In order to manage the data volumes from such a device it is advantageous to choose binary readout for the pixels. In this case, the size of the pixel must be optimised so that only one particle passes though each, otherwise the energy resolution of the device may be degraded. Furthermore, the readout electronics (now integrated into the pixel) must be shielded from the charge-collection regions of the pixel by a deep p-well implant in order to prevent the signal charge drifting to elements of the readout electronics and being lost. Prototype devices (such as the sensor shown in figure 3) have been exposed to the DESY test beam with encouraging results. Further data are being taken in order to fully characterise the sensors.

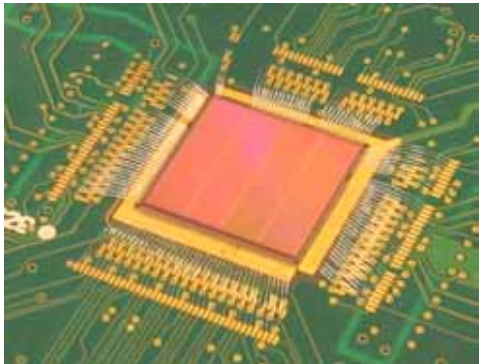

Figure 3: MAPS prototype sensor.

### 2.2. Scintillator Tungsten

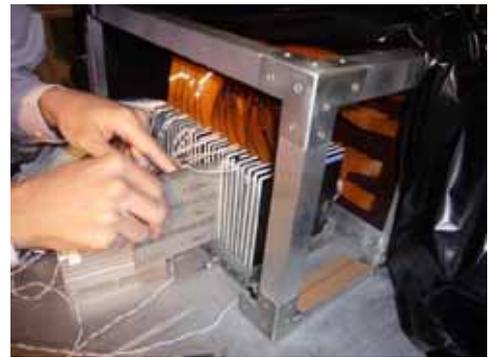

Figure 4: Scintillator Tungsten ECAL prototype.

A complimentary technology for the ECAL is the use of scintillator as the active absorber material. The prototype (shown in figure 4) that was exposed to the DESY test beam has 26 layers using scintillator strips (1 cm wide, 3.5 mm thick). Each strip has a wavelength-shifting fibre that is read-out by a multi-pixel photon counter (MPPC, otherwise known as a silicon photomultiplier, SiPM). These devices are arrays of avalanche photodiodes which give sensitivity to single photons in a very compact silicon package [1]. Three different configurations of fibre and scintillator have been tested in order to determine the optimum configuration. A larger prototype is now under construction. Preliminary results for the en-ergy resolution are shown in figure 5. The stochastic term is $14\%/\sqrt{E}$. Once again, the detector is optimised for shower separation and pattern recognition for particle flow algorithms and therefore does not require very high energy resolution.

## 3. HADRONIC CALORIMETRY

As for the ECAL prototypes, several different technologies are being investigated for hadronic calorimetry (HCAL) at an ILC detector.

### 3.1. Scintillator with Analogue Readout

Figure 6 shows a layer of the HCAL prototype that uses scintillator tiles that are read out by wavelength shifting fibres and SiPMs. The optimal dimensions for the tiles for ILC physics are $3 \times 3$ cm. The SiPM and fibre are integrated into the active volume of the detector. Operation of the prototype has been verified with positrons, with the device showing excellent linearity up to beam energies of 50 GeV and a stochastic term of around $23\%/\sqrt{E}$.

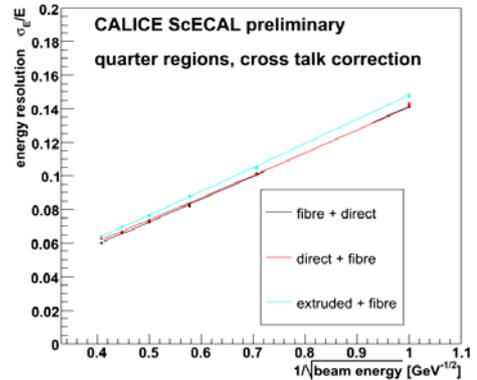

Figure 5: Energy resolution of the scintillator-tungsten ECAL prototype.

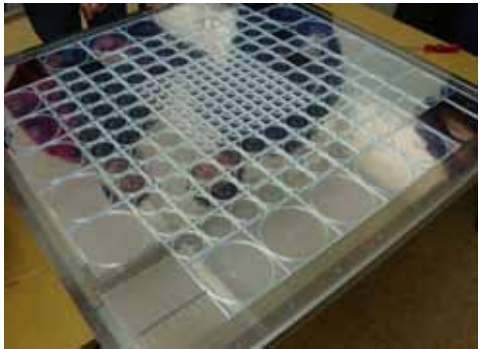

Figure 6: A layer of the analogue scintillator HCAL prototype.

The response of the HCAL prototype to hadrons is shown below in figure 7. The data show excellent Gaussian responses, and studies are now being undertaken to confront the various hadronic interaction models in Geant 4 with the CALICE test beam data.

### 3.2. Digital Readout

Digital technologies trade the relatively small numbers of read out cells in analogue devices for a large number of cells with 1-bit resolutions. Prototypes using resistive plate chambers (RPCs) and gaseous devices with micoMegas, both equipped with digital readout, have been tested. The RPCs, using steel as the absorber, have taken cosmic data and been exposed to the Fermilab test beam (muon, positron, pion and proton beams). The microMegas prototype was exposed to test beam at CERN in July 2008. Figure 8 shows an event in the RPC prototype, whilst figure 9 shows the microMegas digital HCAL prototype in the test beam at CERN. The test beam results from Fermilab have verified the feasibility of the digital HCAL concept using RPCs as the active medium.

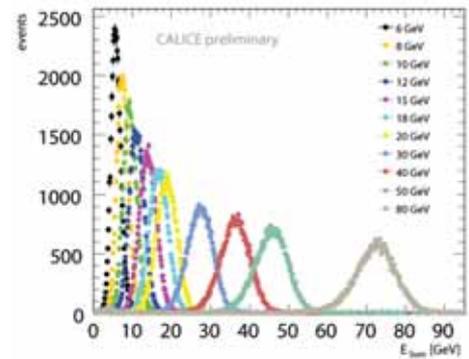

Figure 7: Response of the analogue scintillator HCAL prototype to single pions.

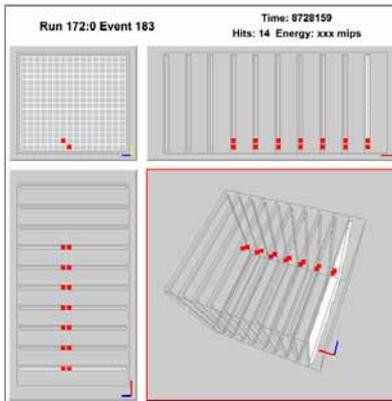

Figure 8: 2-muon event in the RPC digital HCAL at the Fermilab test beam.

## 4. NEXT GENERATION DATA ACQUISITION SYSTEM

The data acquisition (DAQ) system for ILC detectors uses a triggerless design that records all data from every bunch train delivered by the accelerator. Because the event rates are expected to be low (in comparison to the LHC for example), the detectors do not need to be triggered, instead relying on buffering all data in the on-detector electronics before reading out the entire detector.

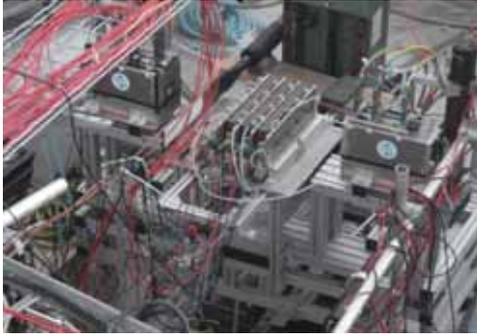

Figure 10: The microMegas digital HCAL prototype in the beam line at CERN in summer 2008.

CALICE is also carrying out R&D into next-generation DAQ architectures using commercial protocols and equipment. The data flow in these systems is packet based, using industry standard protocols (such as Ethernet) as far as possible in order to reduce development risks. The architecture is a classic "funnel" design, relying on detector-specific interfaces (DIF) to the subdetectors which then interface to a common readout architecture for data formatting, selection and storage via a concentrator (LDA). The readout architecture is shown in figure 10. Clock and control is transmitted to the electronics using the same packet-based protocols. Traditional single-ended signals, such as triggers or busies, are relayed using fast commands sent over the same high-speed serial links as the readout data path. All prototyping has been carried out on commercial FPGA development boards, and the serial protocols (including Ethernet), implemented at the FPGA level [2].

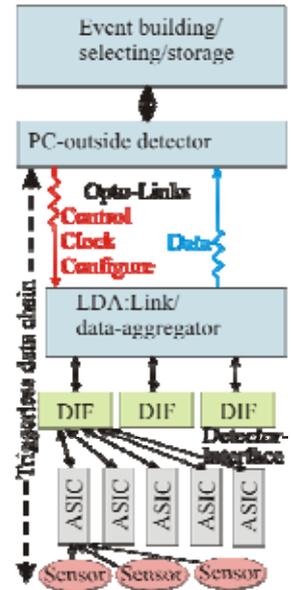

Figure 9: Proposed next-gerneration DAQ architecture.

## 5. CONCLUSIONS

The CALICE collaboration is carrying out R&D into many different technologies for calorimetry at an ILC detector. Many prototypes have successfully taken data in test beams, and these devices are now being used as the basis for large-scale prototypes which are now under construction. These prototypes will implement local data storage in the front-end chips. They will also test the ideas of power pulsing (turning off various parts of the front-end chips) to minimise the power consumption. These two features will enable more realistic tests under ILC-like conditions. Next-generation DAQ architectures are also being developed to read out these prototypes. Together, these systems form the basis for the calorimetry, and potentially the complete readout architecture of a future ILC experiment.